\documentstyle[pra,twocolumn,aps]{revtex}
\RequirePackage{epsfig}
\def\Journal#1#2#3#4{{#1} {\bf #2}, #3 (#4)}
\def\PR{\em Phys. Rev.}
\def\PRL{\em Phys. Rev. Lett.}
\def\PRA{{\em Phys. Rev.} A}
\def\JMP{\em J. Math. Phys.}
\begin{document}
\draft
\title {Comment on ``Strongly interacting one-dimensional Bose-Einstein
condensates in harmonic traps''}
\author{M. D. Girardeau and E. M. Wright}
\address{Optical Sciences Center and Department of Physics,
University of Arizona, Tucson, AZ 85721}
\date{\today}
\maketitle
\begin{abstract}
A recent paper of Tanatar and Erkan
[\Journal{\PRA}{62}{053601}{2000}] discusses a density functional
approach to the impenetrable point Bose gas in one dimension and
an equation for the order parameter of the system, due originally
to Kolomeisky {\it et al.} [\Journal{\PRL}{85}{1146}{2000}], is
derived.  We comment on the regime of validity of such a model and
on qualitative differences between predictions of the density
functional approach and known exact results.
\end{abstract}
The standard Gross-Pitaevskii effective field approach to the
theory of trapped atomic vapor Bose-Einstein condensates (BECs)
has been shown to fail in a regime of very thin waveguides, low
temperatures, low densities, and large positive scattering length
where transverse modes are ``frozen'' and the dynamics reduces to
that of a one-dimensional (1D) gas of impenetrable point bosons
(Tonks gas) \cite{Olshanii,PSW}. A modified effective field
approach has been used recently by Kolomeisky {\it et al.}
\cite{KNSQ} to derive a nonlinear Schr\"{o}dinger equation
specifically adapted to the Tonks gas limit, and their approach
has been further discussed by Tanatar and Erkan \cite{TE}. We
comment herein on the regime of validity of such effective field
models and on important qualitative differences between
predictions of such a density functional approach and known exact
results.

We recently obtained exact many-body solutions for some static and
dynamical properties of trapped 1D Bose gases in the Tonks gas
limit \cite{GW1,GW2,GWT} via appropriate generalizations of our
Fermi-Bose mapping theorem approach \cite{G1,G2}. Detailed
analyses by Olshanii \cite{Olshanii} and by Petrov {\it et al.}
\cite{PSW} show that for sufficiently thin atom waveguides, low
temperatures, and large positive scattering lengths, the dynamics
reduces to that of the Tonks gas (impenetrable point particles) in
the limit of low densities. On the second page of their paper,
Tanatar and Erkan state that ``Our results indicate that a
condensate exists even for an infinitely strongly interacting 1D
model system.'' On the contrary, this is their {\em assumption},
and is in no way {\em indicated} by their results. In fact, it has
been proved rigorously \cite{Lenard,VT} that for an $N$-boson
Tonks gas with no external trapping potential and periodic
boundary conditions, the occupation of the zero-momentum orbital
increases like $\sqrt{N}$ in the thermodynamic limit, not like $N$
as is the case for true BEC. One might hope that a trapped Tonks
gas would show true BEC, but we have recently \cite{GWT} exhibited
the exact many-body ground state of the harmonically trapped Tonks
gas, and our numerical calculation of its reduced one-particle
density matrix shows that the number of atoms in its most highly
occupied orbital increases as $N^{0.59}$ for large $N$.  Thus, in
the limit $N>>1$ one cannot assert that there is a single orbital
that acts as an order parameter or macroscopic wave function for
the whole system and reflects both its density and phase coherence
properties. Thus the order parameter introduced by Kolomeisky {\it
et al.} and Tanatar and Erkan cannot be validly interpreted as a
true condensate wave function. 
There is no true long-range order, so
the interpretation of this quantity as a complex order parameter
is questionable. Kolomeisky {\it et al.} \cite{KNSQ} invoked a
hydrodynamic description such that the magnitude of this quantity
is the square root of the density and its phase is a velocity
potential, which may be expected to capture the long wavelength
behavior but not the strong short-range correlations that prevent
true BEC.

If exact many-body solutions were available for the two-component
Tonks gas, as they are in the one-component case
\cite{GW1,GW2,GWT,G1,G2}, one could compare them with the
approximate two-component nonlinear Schr\"{o}dinger equation
(NLST) solutions of Tanatar and Erkan in order to assess the
reliability of their approach. However, the Fermi-Bose mapping
theorem approach \cite{G1,G2} fails in the two-component case
because for a mixture of two different species of ideal Fermi
gases, the many-body wavefunctions do not vanish at contact of
different species particles, whereas such vanishing is a necessary
condition for validity of the Fermi-Bose mapping theorem. Strong
short-range correlations play an anomalously important role in 1D
(even at low densities) for an obvious topological reason:
Particle collisions are unavoidable in 1D, whereas in higher
dimensions particles can ``detour'' around each other. These
strong short-range correlations (and indeed, {\em all}
two-particle correlations) are omitted in all NLST approaches.

Finally, we comment on some other strengths and weaknesses of the
NLST approach of Kolomeisky {\it et al.} and Tanatar and Erkan. We
note first that comparison of its predictions for the ground state
density profile of a harmonically trapped Tonks gas with the exact
many-body solution \cite{KNSQ,GWT} show that except for small
ripples exhibited by the exact solution, the NLST prediction
agrees closely with the exact solution.  Furthermore, our recent
comparison \cite{GW2} between the NLST prediction and the exact
many-body dynamical solution for a trapped condensate which is
first cut in two by a slowly ramped-up central barrier potential,
then allowed to expand freely, shows that the exact time-dependent
density is closely reproduced by the NLST calculation {\em before}
the expanding clouds collide, but that after their collision the
NLST greatly exaggerates the resultant interference fringes seen
in the exact solution.  The NLST therefore overestimates the phase
coherence present in the system and should not be trusted in
regions where there are significant spatial and/or temporal phase
gradients.
\vspace{0.2cm}

\noindent This work was supported by Office of Naval Research
grant N00014-99-1-0806.


\begin{references}
\bibitem{Olshanii} M. Olshanii, \Journal{\PRL}{81}{938}{1998}.
%
\bibitem{PSW} D.S. Petrov, G.V. Shlyapnikov, and J.T.M. Walraven,
``Regimes of quantum degeneracy in trapped 1D gases'',
cond-mat/0006339 (2000).
%
\bibitem{KNSQ} Eugene B. Kolomeisky {\it et al.},
\Journal{\PRL}{85}{1146}{2000}.
%
\bibitem{TE} B. Tanatar and K. Erkan, \Journal{\PRA}{62}{053601}{2000}.
%
\bibitem{GW1} M.D. Girardeau and E.M. Wright, \Journal{\PRL}{84}{5691}{2000}.
%
\bibitem{GW2} M.D. Girardeau and E.M. Wright, \Journal{\PRL}{84}{5239}{2000}.
%
\bibitem{GWT} M.D. Girardeau, E.M. Wright, and J.M. Triscari, ``Ground state
properties of a one-dimensional condensate of hard core bosons in
a harmonic trap'', cond-mat/0008480 (2000).
%
\bibitem{G1} M. Girardeau, \Journal{\JMP}{1}{516}{1960}.
%
\bibitem{G2} M.D. Girardeau, \Journal{\PR}{139}{B500}{1965}.
%
\bibitem{Lenard} A. Lenard, \Journal{\JMP}{7}{1268}{1966}.
%
\bibitem{VT} H.G. Vaidya and C.A. Tracey, \Journal{\PRL}{42}{3}{1979}.
%
\end{references}
\end{document}